\documentclass[twocolumn,prb,showpacs,floatfix]{revtex4}
\usepackage[dvips]{graphicx}
\usepackage{amsmath,bm}
\usepackage[dvips]{color}
\usepackage{graphicx}
\usepackage{epsfig}

\begin{document}
\title{Strong coupling of a spin qubit to a superconducting stripline cavity}
\author{Xuedong Hu}
\affiliation{Department of Physics, University at Buffalo, SUNY, Buffalo, New York 14260-1500, USA}
\author{Yu-xi Liu}
\affiliation{Institute of Microelectronics, Tsinghua University, Beijing 100084, China}
\affiliation{Tsinghua National Laboratory for Information Science and Technology (TNList), Tsinghua University, Beijing 100084, China}
\affiliation{Advanced Science Institute, RIKEN, Saitama 351-0918, Japan}
\author{Franco Nori}
\affiliation{Advanced Science Institute, RIKEN, Saitama 351-0918, Japan}
\affiliation{Department of Physics, University of Michigan, Ann Arbor, MI 48109-1040, USA}
\begin{abstract}
We study electron-spin-photon coupling in a single-spin double quantum dot embedded in a superconducting stripline cavity.  With an external
magnetic field, we show that either a spin-orbit interaction (for InAs) or an inhomogeneous magnetic field (for Si and GaAs) could produce a
strong spin-photon coupling, with a coupling strength of the order of 1 MHz.  With an isotopically purified Si double dot, which has a very long
spin coherence time for the electron, it is possible to reach the strong-coupling limit between the spin and the cavity photon, as in cavity
quantum electrodynamics. The coupling strength and relaxation rates are calculated based on parameters of existing devices, making this proposal
experimentally feasible.
\end{abstract}
\date{\today}
\maketitle

\section{Introduction}

Confined electron spins in solid-state nanostructures are promising candidates as building blocks of quantum information
processors,\cite{Hanson_RMP, Buluta_RPP, Morton_Nature} with the dual advantages of long quantum coherence times\cite{Bluhm_NP11} and coherent
manipulation of individual qubits.\cite{Petta_Science05, Koppens_PRL08, Morello_Nature10}  The coupling of electron spins can be achieved via
the exchange interaction,\cite{Nowack_Science11} which is short-ranged. Scaling up the spin-qubit architecture invariably requires on-chip
quantum information transfer, whether via moving the electrons themselves\cite{Kane_PRL03, Greentree_PRB04,Taylor_NP05} or via a ``spin
bus''.\cite{Friesen_PRL07} However, electron motion could lead to reduced spin coherence,\cite{Huang_preprint} while the spin bus requires
strong exchange couplings within a spin chain. An enticing alternative to move spin information is to transfer it coherently to a photon, which
is mobile and decoherence-resistant. Such high-fidelity information transfer requires the use of cavity quantum
electrodynamics\cite{Walther_RPP06, Schoelkopf_Review, You_Reviews, Nation_Review, Xiang_Review, Englund_NL10, Frey_preprint} with a microwave
cavity.

The magnetic dipole coupling between a single spin and a photon is very weak, smaller than 1 kHz in a superconducting stripline
resonator,\cite{Schoelkopf_Review} which has a strong vacuum field.  To achieve a strong spin-photon coupling, the electric dipole coupling and
some sort of spin-orbit interaction are required, taking advantage of the so-called electrically-driven spin resonance
(EDSR).\cite{Rashba_SPSS60, Golovach_PRB06, Nowack_Science07, Rashba_PRB08, Sherman_PRB12}  During the past decade, many theoretical proposals
have been put forward to realize coherent spin-photon coupling, including off-resonant Raman scattering off a single
spin,\cite{Imamoglu_PRL99,Childress_PRA04} coupling two-spin states to cavity photons via electric dipole coupling or gate
potential,\cite{Burkard_PRB06,Jin_PRL12} using an InAs nanowire to maximize spin-orbit coupling,\cite{Trif_PRB08} and employing ferromagnetic
leads to create spin-photon coupling.\cite{Cottet_PRL10} However, so far there has been no experimental demonstration of strong interaction
between a single spin and a single photon, so that this remains an important open problem.

Here we propose a spin-photon coupling method with a double quantum dot (DQD) containing a single electron in a superconducting stripline
resonator. The DQD has a large electric-dipole moment, and the spin-electric field coupling is facilitated by either an inhomogeneous magnetic
field (from either a current or a nanomagnet) or spin-orbit interaction (in GaAs and InAs).  We show that the vacuum Rabi frequency of a single
electron spin, under an external magnetic field of 0.1 Tesla, can reach the order of 1 MHz in an inhomogeneous magnetic field, or similar
magnitude in InAs mediated by a strong spin-orbit interaction. We also examine the decoherence of a single spin in a DQD, and identify Si as an
ideal host material because of the absence of nuclear-spin-induced inhomogeneous broadening.  We calculate the spin relaxation rate of a single
electron, and find that it is not a dominant source of decoherence. Combining our results on spin-photon coupling strength and decoherence, we
show that a Si DQD in an inhomogeneous magnetic field is the best structure for an electron spin to reach the strong-coupling limit in
interacting with the photons in a superconducting stripline cavity.

\section{System Setup}


\begin{figure}
\includegraphics[width=3.0in]{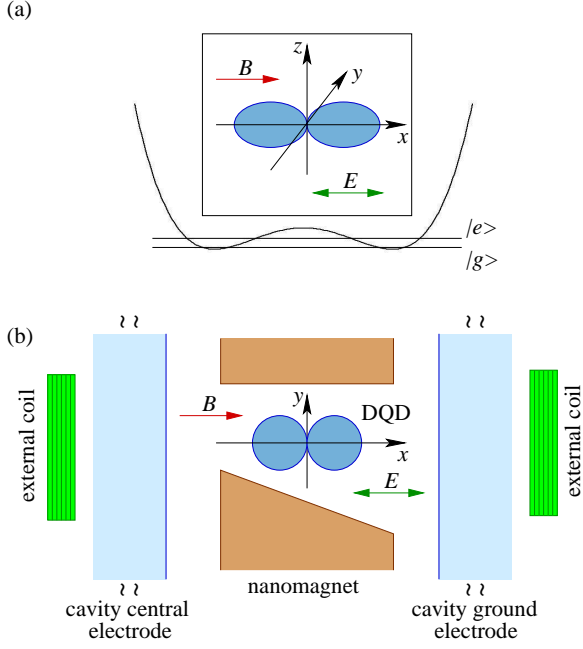}
\caption{ (Color online) Schematics of the two-dimensional double dot in a cavity.  Panel (a) shows that both the cavity electric field and the
external magnetic field are along the interdot axis ($x$-axis).
Panel (b) shows the top view of the combined DQD-nanomagnet-cavity system for the case of a DQD in an inhomogeneous magnetic field.  External
coils are used to produce the uniform magnetic field, while close-by nanomagnets provide the transverse inhomogeneous field.}
\label{fig:schematics}
\end{figure}

The physical system we consider is a semiconductor DQD, with a geometry illustrated in Fig.~\ref{fig:schematics}.  The Hamiltonian of the system
is
\begin{eqnarray}
& & H = H_{\rm DQD} + H_{\rm Zeeman} + H_{\rm SO} + H_{\rm E}\,, \\
& & H_{\rm Zeeman} = \frac{1}{2} g \mu_{\rm B} {\bf B} \cdot {\bf \sigma} \,, \ \ H_{\rm SO} = H_{\rm BR} + H_{\rm I} \,, \nonumber \\
& & H_{\rm BR} = \frac{\alpha_{\rm BR}}{\hbar} \left( \sigma_x P_y - \sigma_y P_x\right) \,, \ \ H_{\rm I} = \alpha_{\rm I} \sigma_y x \,,
\nonumber \\
& & H_{\rm E} = e E x \,. \nonumber
\end{eqnarray}
Here $H_{\rm DQD}$ refers to the orbital part of the single-electron double-dot Hamiltonian, including the confinement potential and the
electron kinetic energy terms.\cite{Hu_PRA00}  To minimize the effect of the external field on the superconducting cavity, we choose an in-plane
uniform magnetic field along the $x$-direction.  This choice also minimizes any effect on the electron orbital motion, so that we only need to
consider the resulting Zeeman splitting.  $H_{\rm SO}$ includes both the normal spin-orbit interaction (such as the Bychkov-Rashba spin-orbit coupling
$H_{\rm BR}$ given here\cite{Rashba_PRB08, footnote1}), and the spin-orbit interaction due to the presence of a spatially-inhomogeneous magnetic
field, such as the one employed in Refs.~\onlinecite{Tokura_PRL06, Pioro-Ladriere_NP08, Obata_PRB10} and given here as $H_{\rm I}$. Lastly,
$H_{\rm E}$ represents the electron interaction with the cavity electric field,\cite{Schoelkopf_Review} which is quantized in terms of the
cavity photon operators $a$ and $a^\dagger$:
\begin{equation}
E = \sqrt{\frac{\hbar \omega_{\rm E}}{Cd^2}} \sin \left(\frac{\pi y}{l}\right) \left( a + a^\dagger \right) \,,
\end{equation}
where $\omega_{\rm E}$ is the photon angular frequency, $C$, $l$, and $d$ are the capacitance, length, and gap width (between the center and the
ground electrodes) of the resonator, respectively. For $\omega_{\rm E} = 10$ GHz and $d = 10 \ \mu$m, the vacuum field is about 0.2
V/m.\cite{Schoelkopf_Review}

\section{Spin-Photon Coupling Strength}

Our focuses are to couple the spin of the single electron to the cavity photons, and to investigate whether we can achieve the strong-coupling
limit in this system.\cite{Devoret_Review}  Here we first project the system Hamiltonian onto the basis states of $|g\rangle |\!\uparrow
\rangle$, $|g\rangle |\!\downarrow \rangle$, $|e\rangle |\!\uparrow \rangle$, and $|e\rangle |\!\downarrow \rangle$, where $|g\rangle = \alpha
|L\rangle + \beta|R\rangle$ and $|e\rangle = \beta |L\rangle - \alpha|R\rangle$ are the ground and first excited orbital states of the
tunnel-coupled DQD with energies $\epsilon_g$ and $\epsilon_e$.  $|L\rangle$ ($|R\rangle$) is the single-dot ground orbital state in the left
(right) dot, while $|\!\uparrow \rangle$ and $|\!\downarrow \rangle$ are the spin eigenstates of $H_{\rm Zeeman}$.  The inter-dot tunnel
coupling is $\langle L|H|R\rangle = t$.  A straightforward inspection of the matrix elements of the total Hamiltonian shows that the largest
spin-photon coupling is achieved when there is no bias between the two dots, when $\epsilon_0 = \epsilon_e - \epsilon_g = 2|t|$, which we choose
at 40 $\mu$eV, or about 10 GHz. This energy is much smaller than the single-dot excitation energy of $\sim 1$ meV, justifying our focus on this
sub-Hilbert space for our calculations. The total Hamiltonian now becomes
\begin{widetext}
\begin{equation}
H = \left(
\begin{array}{cccc}
\epsilon_g - \epsilon_{\rm Z} & 0 & -eEL &  -\lambda_x + i \alpha_{\rm I} L \\
0 & \epsilon_g + \epsilon_{\rm Z} & \lambda_x - i \alpha_{\rm I} L & -eEL \\
-eEL & \lambda_x + i \alpha_{\rm I} L & \epsilon_e - \epsilon_{\rm Z} & 0 \\
 -\lambda_x - i \alpha_{\rm I} L & -eEL & 0 & \epsilon_e + \epsilon_{\rm Z}
\end{array}
\right)\,.
\end{equation}
\end{widetext}
Here $\epsilon_Z = \frac{1}{2} g\mu_B B$ is the Zeeman energy due to the uniform in-plane magnetic field, $E$ is the electric field operator,
and $L$ is the half-interdot-distance for the DQD.  The matrix element for the inhomogeneous field induced spin-orbit coupling is quite
simple ($i \alpha_{\rm I} L$).  The physical picture is straightforward as well: the field inhomogeneity leads to different quantization
axes for the two dots, so that a spin eigenstate in one dot is a superpositioned state in the other.  When driven by an electric field,
this spin mixture leads to an AC transverse magnetic field, which in turn produces EDSR.\cite{Tokura_PRL06}
The ``intrinsic'' (by which we mean that the interaction is a material/structure property, and is not due to an applied field or magnet)
spin-orbit coupling parameter is:
$$\lambda_x = \frac{\alpha_{\rm BR} LS}{a^2
\sqrt{1-S^2}} \,,$$
where
$$S = \langle L|R\rangle = e^{-(L/a)^2}$$ 
is the interdot wave function overlap, and
$a$ is the radius of the Gaussian single-dot ground state wave function.  Here $\lambda_x$ originates from the $\sigma_y P_x$ term in $H_{\rm
BR}$. It is related to the $x$-direction electric dipole moment of the double dot, and is the main driving force behind the EDSR in the presence
of intrinsic spin-orbit interaction.
%

When $\epsilon_{\rm Z} \ll \epsilon_0$, we can focus on the spin dynamics by projecting our system Hamiltonian to the sub-Hilbert-space spanned
by the ground orbital state and the spin eigenstates $|g\rangle \otimes |\!\uparrow \rangle$ and $|g\rangle \otimes |\!\downarrow \rangle$.
According to L\"{o}wdin's perturbation theory,\cite{Winkler_book} the matrix elements of an effective Hamiltonian (at the lowest order) in the
reduced Hilbert space take the form
\begin{equation}
H_{ij}^{\prime} = H_{ij} + \frac{1}{2} \sum_k \left\{ \frac{H_{ik} H_{kj}}{H_{ii} - H_{kk}} + \frac{H_{ik} H_{kj}}{H_{jj} - H_{kk}} \right\} \,,
\end{equation}
where the indices $i$ and $j$ refer to states in the targeted sub-Hilbert space, while $k$ refers to states outside this subspace.  The
off-diagonal term that leads to the cavity-electric-field-driven spin rotation thus takes the form
\begin{eqnarray}
H^{\prime}_{12} 
& = & - 2 e E L \left( \frac{i \alpha_{\rm I} L}{\epsilon_0} + \frac{\lambda_x \epsilon_{\rm Z}}{\epsilon_0^2} \right)  \label{eq:H12}\,.
\end{eqnarray}
This spin-cavity coupling term has several interesting features.  First, the denominators for the terms in Eq.~(\ref{eq:H12}) are $\epsilon_0$
or $\epsilon_0^2$.  For a single circular dot (SQD) these terms would have been $\hbar \omega_0$ or $(\hbar \omega_0)^2$, where $\hbar \omega_0$
is the single-particle excitation energy, on the order of 1 meV in GaAs/Si and 10 meV in InAs.  Since $\epsilon_0 \ll \hbar \omega_0$ in a
double dot, the spin-photon coupling strength tends to be stronger in a DQD compared to a SQD, even after the reduction due to tunnel coupling
is included.  Alternatively, one can use an elongated dot to enhance the spin-photon coupling as well, with InAs nanowires as a prime
example.\cite{Trif_PRB08}  Here the longitudinal confinement is weaker, so that the enhancement due to the stronger orbital coupling, like that
in a DQD, can be achieved as well.  Another feature of Eq.~(\ref{eq:H12}) is that the spin-orbit-induced spin rotation term contains an extra
factor of $\epsilon_{\rm Z}/\epsilon_0$ as compared to the inhomogeneous-field-induced spin rotation.  This extra factor originates from the
breaking of the Kramers degeneracy by an external magnetic field, since spin-orbit interaction itself does not break the time-reversal symmetry, while an inhomogeneous
magnetic field does.

In Table~\ref{table:matrix_elements} we present our results on both DQD and SQD in terms of the effective spin coupling matrix element.  For our
SQD calculations we included the ground and the first two excited orbitals.  Our choices of half-interdot-distance $L$ for DQDs are determined
by the assumption that a single-dot confinement energy in GaAs or Si is about 1 meV, and the interdot tunnel coupling should be about 40
$\mu$eV. While the conduction electron effective mass in GaAs is smaller, leading to larger $L$, the $g$-factor in GaAs is small ($\sim$0.4), so
that under the same nanomagnet the coupling strengths in GaAs and Si are similar. Results for an elongated single dot are also included (from
Refs.~\onlinecite{Trif_PRB08,Borhani_PRB12}) for comparison.

\begin{table}[h]
\begin{tabular}{|c|c|c|}
\hline
& $H_{12}$ & interaction strength \\ \hline
GaAs DQD & $-2ieEL^2 \alpha_I\epsilon_0^{-1}$ & 0.5 MHz \\
inhomogeneous field & & ($\alpha_I = 1$ Tesla/$\mu$m) \\ \hline
Si DQD & $-2ieEL^2 \alpha_I\epsilon_0^{-1}$ & 0.6 MHz \\
inhomogeneous field & & ($\alpha_I = 1$ Tesla/$\mu$m) \\ \hline
DQD spin-orbit & $- 2eEL \lambda_x \epsilon_{\rm Z}\epsilon_0^{-2}$ &  50 kHz (GaAs) \\
& & to 1 MHz (InAs) \\
\hline
SQD & $-ieEa^2 \alpha_I E_0^{-1}$  & 0.2 kHz (InAs) \\
inhomogeneous field & & to 5 kHz (Si) \\ \hline
SQD spin-orbit & $-eE\alpha_{\rm BR} \epsilon_{\rm Z}E_0^{-2}$ & 10 kHz to 0.1 MHz \\ \hline
Elongated SQD SO & $-eE\alpha_{\rm BR} \epsilon_{\rm Z} E_x^{-2}$ & 100 kHz to 1 MHz \\ \hline
\end{tabular}
\caption{Spin-photon coupling strength in single and double dots.  The estimated magnitudes are obtained by assuming a double dot with half
interdot distance $L = 60$ nm in GaAs and 35 nm in Si, $\epsilon_0 = 40 \mu$eV, and a vacuum electric field of 0.4 V/m.  The single dot
parameters are $E_0 = 1$ meV for a GaAs dot and 30 meV for an InAs dot.  For elongated dots, we choose $E_x = 0.1$ meV for GaAs and 1 meV for
InAs.} \label{table:matrix_elements}
\end{table}

\section{Strong Coupling Limit: Spin Relaxation}

To achieve the strong-coupling limit, which would allow coherent information transfer between the spin and the cavity photon, the spin-photon
coupling strength needs to satisfy $|H_{12}| > \hbar \gamma$, where $\gamma$ is the total decoherence rate of the spin-cavity system, including
cavity loss and spin decoherence. With a quality factor of $Q > 10^5$ and a photon frequency of 5 GHz, \cite{Johnson_NP10} the photon decay rate
is $<0.05$ MHz.  Thus photon loss should not pose any significant problem, at least not to demonstration-of-principle type experiments. Spin
decoherence, on the other hand, is a more serious issue.  While the true coherence times of spin qubits are generally long (A $T_2$ of the order
of 200 $\mu$s has recently been measured in a GaAs DQD\cite{Bluhm_NP11}), the environmental nuclear spins generally cause significant
inhomogeneous broadening in III-V quantum dots.  For example, the dephasing time due to inhomogeneous broadening from nuclear spins in GaAs is
$T^*_2 \sim 10$ ns.\cite{Petta_Science05} While this dephasing mechanism does not correspond to true decoherence, it does imply a spin frequency
uncertainty on the order of 100 MHz at any particular time.  With the cavity frequency fixed over time, such a spin frequency shift would make
spin-cavity resonance untenable. Without a viable solution to this dephasing problem,\cite{footnote2,Foletti_NP09,Reilly_Science08} it would be
impossible to achieve strong spin-photon coupling in a GaAs SQD or DQD.  While InAs does have a much stronger spin-orbit coupling, the
spin-photon coupling strength is still not of the same order as the dephasing rate (see Table I), making the strong-coupling limit difficult to
reach as well.

A more promising candidate to achieve strong spin-photon coupling via a stripline resonator may be an isotopically-purified Si DQD. While
spin-orbit coupling is small in Si (with a coupling strength about one tenth that of GaAs), the spin coherence time is very long, especially in
isotopically purified $^{28}$Si samples, where the nuclear-spin-induced dephasing can be $< 0.1$ MHz,\cite{Assali_PRB11} and the true
decoherence rate is much smaller.\cite{Morton_Nature} Combining this long decoherence time with an external inhomogeneous magnetic field,
achieving the strong-coupling limit becomes a more realistic goal.

\begin{figure}
\includegraphics[width=2.8in]{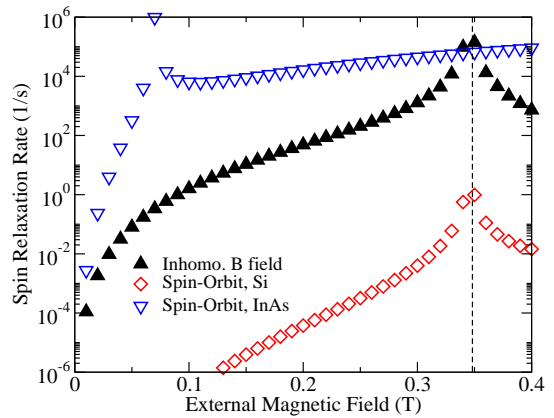}
\caption{ (Color online) Single-spin relaxation rate in a Si or InAs DQD as a function of the applied field.  Here the tunnel splitting of the
double dot orbital state is $\epsilon_0 = 40 \ \mu$eV (the vertical line represents the resonant condition when $2\epsilon_Z = \epsilon_0$ for
Si, with a corresponding magnetic field of 0.345 T), and the interdot distance is 100 nm for Si and 50 nm for InAs. The black filled triangles
represent the spin relaxation due to spin mixing from the inhomogeneous magnetic field along the interdot ($x$) direction, while the red
unfilled diamond (for InAs) and blue unfilled triangles (for Si) are due to Rashba spin-orbit coupling.  The spin relaxation rate for InAs
peaked at a different magnetic field because of the very different $g$-factor (we used $g=10$ here).} \label{fig:spin_relaxation}
\end{figure}

With pure dephasing due to nuclear spins not a significant problem in a Si DQD, we still have to clarify the spin relaxation rates there. While
it is known that spin relaxation in an SQD is extremely slow,\cite{Golovach_PRB06,Jiang_PRL10} past calculations have also shown that for a
single electron in a DQD, there could exist spin hot spots when the electron Zeeman energy and the DQD tunnel splitting is on
resonance.\cite{Stano_PRB05} We therefore calculate the single-spin relaxation rate in a Si DQD due to the presence of spin-orbit interaction
and inhomogeneous magnetic field. The results are presented in Fig.~\ref{fig:spin_relaxation}.  The peak in the relaxation rate occurs when
$|g\uparrow\rangle$ and $|e\downarrow\rangle$ are degenerate, {\it i.e.} $\epsilon_0 = 2 \epsilon_Z$.  With $\epsilon_0 = 40 \ \mu$eV, the
resonant magnetic field is 0.345 T for Si. Even for $B = 0.3$ Tesla, which is quite close to the hot spot, the spin relaxation rate is only
$\sim 1$ kHz, posing no problem to the spin-photon coupling scheme.

\section{Discussions and Conclusions}

Achieving strong spin-photon coupling would not only allow long-distance quantum communication for spin qubits, but also open a range of new
possibilities in spin and photon physics, including spin and photon manipulations, and dispersive spin measurement.\cite{Schoelkopf_Review}  The
latter could present a viable alternative to the charge sensor and spin-blockade-based measurement technique\cite{Hanson_RMP, Petta_Science05}
that is widely used today for spin qubits. Furthermore, a hybrid design of qubit architecture, with a superconducting stripline cavity as the
backbone, and involving various kinds of qubits (such as atomic, ionic, superconducting, and spin qubits) would also become
feasible.\cite{Xiang_Review}

As we discussed above, a spin-photon coupling driven by an inhomogeneous magnetic field in Si is probably the best combination to achieve the
strong-coupling limit.  We note that the strength of this EDSR scheme originates from its ``directness'', in the sense that the electron spin is
directly rotated by the non-uniform magnetic field, as opposed to the case with spin-orbit interaction, which relies on an applied magnetic
field to lift the spin degeneracy.  For a spin-orbit qubit, such as what is discussed in Refs.~\onlinecite{Nadj-Perge2010} and
\onlinecite{Schroer_PRL11}, the effects of the spin-orbit interaction is already incorporated in the form of a renormalized anisotropic and
confinement-dependent $g$-factor.  For such a system, a uniform applied field would produce an apparent inhomogeneous Zeeman splitting across a
DQD, so that the EDSR for the qubit can be considered as an EDSR in an inhomogeneous magnetic field.  It is also worth noting here that an
inhomogeneous magnetic field is not only useful for spin manipulation, but could also lead to an alternative approach for spin measurement, as
was discussed in Ref.~\onlinecite{TahanThesis}.  In other words, a combined DQD-cavity-nanomagnet system has the versatility for studying
several different aspects of the coupled spin-photon dynamics.

The main drawback of using the inhomogeneous magnetic field is the added device complexity, with an additional metallic layer in the system and
a magnetic material close to the superconducting electrodes of the cavity.  To enhance the electric-field-DQD coupling, one could connect the
DQD plunger gates directly to the cavity electrodes.\cite{Jin_PRL12}  Such a configuration should allow a larger gap between the cavity
electrodes, so that the nanomagnet can be more easily incorporated.

Compared to the schemes of two-spin coupling to cavity photons,\cite{Burkard_PRB06} our single-spin approach avoids the Coulomb interaction
between the two electrons, which works to reduce the spin-photon coupling strength.  Furthermore, maximizing spin-photon (or any other two-level
system-to-photon) coupling strength requires a strong vacuum electric field, which can be achieved with a narrower gap for the cavity. However,
such reduction in device dimension has to be carefully managed so that it does not conflict with the devices (such as gated quantum dots and
nanomagnets) that are incorporated into the cavity.

The spin-orbit-based spin-photon coupling mechanism here is similar to the system discussed in Ref.~\onlinecite{Trif_PRB08}, with both
approaches achieving an enhancement of the spin-photon coupling by reducing the orbital state gap, whether through tunnel coupling or by having
an elongated quantum dot. The DQD configuration has a clean low-energy spectrum that allows a simpler and more transparent mathematical
treatment, and an elongated dot could easily become a DQD due to local disorder, but a DQD has to overcome suppression by tunnel coupling. The
spin-photon coupling strengths in these methods are similar after all the factors are considered, as shown in Table I.  If a frequency locking
mechanism is ever discovered to overcoming the inhomogeneous broadening in the III-V quantum dots, an InAs DQD or elongated dot with a single
electron\cite{Petersson_preprint} would be a good candidate to achieve strong spin-photon coupling as well.

In conclusion, we propose to achieve the strong-coupling limit between an electron spin qubit and a cavity photon mode by using a single-spin
double quantum dot and a superconducting stripline cavity. We show that an inhomogeneous magnetic field could lead to a strong interaction
between the spin and the cavity photon. We estimate the coupling strength and relaxation rates based on existing devices and technology, so that
the coupling method proposed here should be feasible with a single-electron Si double dot.  We also show that in materials such as InAs, the
inhomogeneous broadening due to nuclear spins needs to be overcome to achieve strong coupling between electron spins and photons as well.

We acknowledge useful discussions with Alexander Khaetskii, Jason Petta, Karl Petersson, Steve Lyon, Seigo Tarucha, and Charles Tahan.  X.H. thanks support by US ARO (W911NF0910393), DARPA QuEST through AFOSR, and NSF PIF (PHY-1104672). Y.X.L. was supported by NSFC under Grant Nos.~10975080 and 61025022. F.N. was partially supported by the ARO, JSPS-RFBR contract No.~12-02-92100, Grant-in-Aid for Scientific Research (S), MEXT Kakenhi on Quantum Cybernetics, and the JSPS via its FIRST program.


\begin{thebibliography}{99}
\bibitem{footnote1} While our calculations are explicitly for Rashba spin-orbit interaction, including Dresselhaus type of spin-orbit
interaction will not change the qualitative discussions.
\bibitem{footnote2} It has been shown that feedback mechanisms allow a buildup and stablization of a finite nuclear field gradient between two
quantum dots,\cite{Foletti_NP09} while narrower inhomogeneous broadening (down to 10 MHz, instead of 100 MHz) may be possible as well with such
feedback control.\cite{Reilly_Science08}  However, it is not clear at present whether such techniques can be pushed to achieve another two
orders of magnitude suppression of inhomogeneous broadening.
\end{thebibliography}
\end{document}